\documentclass[12pt]{iopart}
\usepackage{graphicx}
\usepackage{xcolor}

\begin{document}
\pdfminorversion=4 

\title[Invisible $Z$-boson ratio and asymmetry]{Predictions of the ratio and asymmetry probes of the invisible $Z$-boson decay}

\author{Kadir Saygin}

\address{Faculty of Aviation and Space Sciences, Necmettin Erbakan University, Konya, Turkey}
\ead{kadir.ocalan@erbakan.edu.tr}
\vspace{10pt}
\begin{indented}
\item[]April 2023
\end{indented}

\begin{abstract}
Higher-order predictions through the combined accuracy including next-to-leading order (NLO) electroweak (EW) and next-to-NLO (NNLO) quantum chromodynamics (QCD) corrections in underlying perturbation theories are presented thoroughly for the invisible decay of the $Z$ boson into neutrino pair relative to its decay into charged-lepton pair (leptonic decay). The combined NNLO QCD+NLO EW predictions are achieved based on the fully-differential calculations of cross sections of both the invisible and leptonic processes in proton-proton ($pp$) collisions at 13 TeV center-of-mass energy. Differential distributions of cross-section ratios of the invisible process to the leptonic process are presented as a function of the transverse momentum of the $Z$ boson $p^{Z}_{\rm{T}}$. For the first time, the predictions for differential distributions of cross-section asymmetries between the invisible process and the leptonic process are presented in bins of the $p^{Z}_{\rm{T}}$. The cross-section ratio and asymmetry distributions, which are referred to as the invisible probes, are considered to be important for controlling the invisible process by the leptonic process of the $Z$ boson and probing deviation from the Standard Model (SM) for new-physics searches. The predictions are extensively presented beyond the $Z$-boson mass resonance region to assess the potential of the invisible ratio and asymmetry probes for new-phenomena searches in high-invariant mass region of the lepton-pair final states. Various tests with threshold requirements of transverse momenta of neutrino pair and leptons are performed to assess the impact on the combined predictions. The invisible ratio and asymmetry probes are proposed to be important probes for indirect searches of new-physics scenarios.        

\end{abstract}

\vspace{2pc}
\vspace{2pc}

\textcolor{red}{This is the version of the article before peer review or editing, as submitted by the author to the \textit{Physica Scripta}. IOP Publishing Ltd is not responsible for any errors or omissions in this version of the manuscript or any version derived from it. The Version of Record is available online at \textbf{https://doi.org/10.1088/1402-4896/ace80a}}

%
%
\maketitle
%
%

\section{Introduction}
\label{int}
Electroweak gauge boson production mechanisms including those of $W$ and $Z$ bosons represent major benchmark processes of the Standard Model (SM) of high-energy particle physics. Subsequent leptonic decay channels of these bosons have sizable production cross sections without large residual contributions from background processes at a hadron collider, presently at the CERN Large Hadron Collider (LHC). Their leptonic decays enable high-precision measurements leading to several important tests of both quantum chromodynamics (QCD) and electroweak (EW) sectors of the SM. Particularly leptonic decays of the $Z$ boson $Z \rightarrow l^{+}l^{-}$, referring to neutral-current mechanism of Drell-Yan lepton-pair production in the $Z$-boson mass resonance region, have been under precise scrutinization by the LHC experiments using proton-proton ($pp$) collisions data by means of both inclusive and differential cross-section measurements, such as in the Refs.~\cite{ATLAS:2016fij,ATLAS:2019zci,CMS:2018mdl,CMS:2019raw,LHCb:2021huf}. These measurements constitute valuable inputs for improving description of parton distribution functions (PDFs) in the proton and facilitate stringent tests of theoretical models and calculations of the SM. The field of theoretical calculations has advanced accordingly to standardize inclusion of higher-order corrections at next-to-leading order (NLO) in EW~\cite{Kuhn:2005az,Denner:2011vu,Dittmaier:2014qza,Lindert:2017olm} and next-to-NLO (NNLO) in QCD~\cite{Hamberg:1990np,Melnikov:2006kv,Catani:2009sm} perturbation theories for precise and accurate predictions of the $Z$-boson decay processes.                

In the searches beyond the SM, $Z$ boson decays represent major background processes for various new-physics models (e.g. dark matter and supersymmetry) in final states with energetic jets and large missing transverse momentum~\cite{CMS:2014jvv, ATLAS:2017bf,CMS:2017zts,ATLAS:2018nda,CMS:2019zmd,ATLAS:2021kxv,CMS:2021far}. Missing transverse momentum corresponds to a momentum imbalance in transverse plane of detectors relative to the $pp$ beam direction, represented by $p^{\rm{miss}}_{\rm{T}}$, stemming from presence of particles that do not involve in either strong or electromagnetic interactions with detector components. These particles are referred to as $\emph{invisible}$ particles, such as weakly-interacting neutrino pair from the invisible decay of the $Z$ boson $Z \rightarrow \nu \bar{\nu}$. The invisible $Z$-boson decay can be translated into a strong probe of new-physics searches, where such invisible signatures might also indicate a pair of weakly-interacting massive particles (WIMPs) $\chi \bar{\chi}$ as dark matter candidates~\cite{Abdallah:2015ter} that could show up at the LHC. Invisible decay of the Higgs boson $H \rightarrow \nu \bar{\nu}$ through vector-boson fusion (VBF) is another important production mechanism for searches of weakly-interacting new particles, where the dominant SM process leading to the same final states is the invisible decay of the $Z$ boson in association with jets. The LHC experiments have previously set some limits regarding new-physics searches for the Higgs boson decaying invisibly via VBF~\cite{ATLAS:2018bnv,CMS:2018yfx}. Consequently, precise experimental and theoretical control of the invisible $Z$-boson decay is crucial in the $p^{\rm{miss}}_{\rm{T}}$(+jets) final states to enhance the potential of searches for new phenomena. The differential cross-section measurement for the transverse momentum of the invisibly decaying $Z$ boson has been reported recently by the CMS Collaboration based on $pp$ collisions data at 13 TeV center-of-mass energy~\cite{CMS:2020hkd}. The ATLAS Collaboration reported measurements for the ratio of $Z \rightarrow \nu \bar{\nu}$/$Z \rightarrow l^{+}l^{-}$~\cite{ATLAS:2017txd} and for the $Z \rightarrow \nu \bar{\nu}l^{+}l^{-}$ production~\cite{ATLAS:2019xh} at 13 TeV $pp$ collisions energy. These measurements methodologically exploited the similarity in kinematic characteristics between the $Z$-boson decay into neutrino pair and its decay into well-measured charged lepton pair (electron-positron or muon-antimuon pair). On the theory side, high-order calculations are required in both QCD and EW perturbation domains to predict precisely cross section of the invisible $Z$-boson decay. Theoretical controls of the invisible decay can be improved by making use of the leptonic $Z$-boson decay in support of probing new physics. Any deviation from high-precision calculations in comparison with experimental results would reveal reliable evidence of new-physics models.       

This paper presents genuine higher-order theoretical predictions at 13 TeV center-of-mass energy for the invisible process $pp \rightarrow Z \rightarrow \nu \bar{\nu}$ in relation with the leptonic process $pp \rightarrow Z \rightarrow l^{+}l^{-}$ at NNLO and NLO accuracies in QCD and EW perturbation theories, respectively. Differential distributions of cross-section ratios of the invisible process to the leptonic process are presented extensively through the combined NNLO QCD+NLO EW predictions as a function of the transverse momentum of the $Z$ boson $p^{Z}_{\rm{T}}$. Differential ratio distributions are valuable for new-physics searches as they are sensitive to probe any deviation in the $Z$-boson branching fraction or decay width to neutrino pair. For the first time, differential distributions of cross-section asymmetries between the invisible process and the leptonic process as a function of the $p^{Z}_{\rm{T}}$ are presented based on the combined NNLO QCD+NLO EW predictions. Differential asymmetry distributions render the invisible process a more powerful probe for both searches in $p^{\rm{miss}}_{\rm{T}}$(+jets) final states and searches in Drell-Yan lepton-pair final states in high-invariant mass $m_{ll}$ region. Thereby the predicted results of this paper unveil differential ratio and asymmetry distributions as important invisible probes by means of both enabling precise controls of the invisible process and contributing to extension of indirect new-physics searches beyond the SM.   

\section{Methodology}
\label{meth}
Differential cross-section calculations for the invisible and leptonic decays of the $Z$ boson are carried out using the computational framework MATRIX (version 2.1.0.beta2)~\cite{Catani:2009sm,Grazzini:2017mhc,Grazzini:2019jkl}. The Catani-Seymour dipole-subtraction method is exploited by default within the MATRIX framework to perform fixed-order calculations at NLO QCD~\cite{Catani:1996jh,Catani:1996vz} and NLO EW~\cite{Dittmaier:1999mb,Dittmaier:2008md,Gehrmann:2010ry,Kallweit:2017khh,Schonherr:2017qcj}. NNLO QCD calculations are achieved using the well-established \emph{$q_{\rm{T}}$}-subtraction formalism~\cite{Catani:2007vq,Catani:2012qa}. A cut-off value for the slicing parameter $r_{\rm{cut}}$ of the \emph{$q_{\rm{T}}$}-subtraction method is employed as $r_{\rm{cut}}=0.0015 (0.15\%)$ for proper regularization of soft and collinear divergent terms of the real radiation in the underlying perturbative expansion. The MATRIX incorporates the OpenLoops 2 program to facilitate computations of tree-level and one-loop scattering amplitudes~\cite{Matsuura:1988sm,Cascioli:2011va,Denner:2016kdg,Buccioni:2017yxi,Buccioni:2019sur}, as well as color- and spin-correlated amplitudes for the elimination of divergences. PDF information is necessarily included in the calculations based on the NNPDF31~\cite{Bertone:2017bme} PDF models utilizing the LUXqed methodology~\cite{Manohar:2016nzj}, which is required to account for quantum electrodynamics (QED) effects in determination of photon density. More specifically, NNPDF31\_nnlo\_as\_0118\_luxqed PDF set is used at both NLO and NNLO perturbative accuracies based on the strong coupling constant $\alpha_{S}(m_{Z})=$ 0.118 value. The LHAPDF (v6.4.0) platform~\cite{Buckley:2014ana} is exploited for evaluation of the PDF sets in the computations. In the combined NNLO QCD+NLO EW calculations, a standard additive prescription from the Ref.~\cite{Grazzini:2019jkl} is used to combine NNLO QCD and NLO EW corrections. Hereby the differential cross section at NNLO QCD+NLO EW can literally be expressed in terms of corresponding correction factors $\delta_{\rm{NNLO~QCD}}$ and $\delta_{\rm{NLO~EW}}$ relative to the leading-order (LO) differential cross section $d\sigma_{\rm{LO}}$ as the following

\begin{equation}
\label{eqn:1}   
\begin{array}{l}
\displaystyle
d\sigma_{\rm{NNLO~QCD}+\rm{NLO~EW}}=d\sigma_{\rm{LO}}(1+\delta_{\rm{NNLO~QCD}}+\delta_{\rm{NLO~EW}}).
\end{array}
\end{equation} 

The $Z$ boson is treated off-shell in both the invisible and leptonic processes. The $Z$-boson mass $m_{Z}$ is defined by the so-called complex-mass scheme~\cite{Denner:2005fg} which is based on the definition of EW mixing angle as $\sin^{2}\theta_{W}$ (and $\alpha$) in terms of complex $W$- and $Z$-boson mass terms. Input EW parameters are supplemented in the computations according to the $G_{\mu}$ input scheme. The input values of this scheme are used the same as the PDG values~\cite{ParticleDataGroup:2016lqr} for the $m_{Z}$, the decay width $\Gamma_{Z}$, and the Fermi constant $G_{\rm{F}}$ as the followings

\begin{equation}
\label{eqn:2}   
\begin{array}{l}
\displaystyle
\it{m_{Z}}=\rm{91.1876}~\rm{GeV}, \Gamma_{\it{Z}}=2.4952 ~ \rm{GeV}, \rm{and} ~ \it{G_{\rm{F}}}=\rm{1.16639}\! \times\! \rm{10^{-5}} ~ \rm{GeV}^{-2}. \\
\end{array}
\end{equation}    
The baseline fiducial phase space selection at particle level is adopted from the CMS measurement of differential cross section of the invisible $Z$-boson decay at 13 TeV~\cite{CMS:2020hkd}. Since neutrinos leave detector system undetected, invisible $Z$-boson decays can only be identified at sufficiently large $p^{Z}_{\rm{T}}$ values, which is translated into significant amount of missing transverse momentum in the decay, i.e., large $p^{\rm{miss}}_{\rm{T}}$ values. The transverse momentum of the neutrino pair is therefore required to be $p^{\rm{miss}}_{\rm{T}}>$ 200 GeV in the baseline fiducial selection. No geometrical acceptance requirement in terms of pseudorapidity is applied for the invisible process. In the leptonic process two same-flavour leptons (either electrons or muons) with opposite electric charge signs are required in the final state without any requirements either on the lepton pseudorapidity or on the lepton transverse momentum, i.e., $p^{l}_{\rm{T}}>$ 0. Nevertheless, an invariant mass $m_{ll}$ window requirement $76<m_{ll}<106$ GeV for lepton pair emitted in final states is imposed to minimize contribution from the $\gamma^{*} \rightarrow l^{+}l^{-}$ process and its interference with the leptonic $Z$-boson process. In both invisible and leptonic processes, final states are allowed to be accompanied by emission of at least one recoiling jet which is accurate up to NNLO in the perturbative QCD calculations. No requirements for the jet transverse momentum and pseudorapidity are imposed at all. Furthermore, photon recombination is switched on in the computations to account for photon-induced effects from quark- photon decay channels at NLO EW. Leptons are treated with dressed lepton definition throughout the entire calculations, where collinear photon and charged leptons are recombined into dressed leptons within a cone of radius $\Delta R=$0.1. Leptons are treated 'dressed' based on this dressed lepton definition in a consistent manner throughout the entire paper.    

Differential cross-section calculations for both the invisible and leptonic processes are achieved at fixed-order (N)NLO and at combined NNLO QCD+NLO EW perturbative accuracies as a function of the $p^{Z}_{\rm{T}}$. Similar to the ATLAS measurement at 13 TeV~\cite{ATLAS:2017txd}, a cross-section ratio observable $R_{\rm{inv}}$ can be exploited, which is defined in this paper to be fully differential in the $p^{Z}_{\rm{T}}$ as

\begin{equation}
\label{eqn:3}   
R_{\rm{inv}}=\frac{d\sigma(Z \rightarrow \nu \bar{\nu})/dp^{Z}_{\rm{T}}}{d\sigma(Z \rightarrow l^{+}l^{-})/dp^{Z}_{\rm{T}}}.
\end{equation}
The invisible ratio $R_{\rm{inv}}$ can be used to control the invisible process relative to the experimentally well-known leptonic $Z$-boson decay, as well as to probe any departure from the SM predictions which would indicate signs for new-physics phenomena. More specifically an unexpected deviation of the $R_{\rm{inv}}$ in the form of an excess in the branching fraction or decay width of the invisible $Z$ boson would reveal signs of new particles emitted in the final state. In addition, a cross-section asymmetry observable $A_{\rm{inv}}$ in this context is introduced to enable more powerful controls of the invisible process by the leptonic process. The invisible asymmetry $A_{\rm{inv}}$ is defined analogously with the definition of lepton charge asymmetry of the W boson~\cite{Catani:2010en}, for which neutrino-pair and lepton-pair final states are regarded as two distinct parameter spaces of the Z-boson decay with similar kinematics. The $A_{\rm{inv}}$ is thus defined in terms of differential cross sections as a function of the $p^{Z}_{\rm{T}}$ by the following expression 
    
\begin{equation}
\label{eqn:4}   
A_{\rm{inv}}=\frac{d\sigma(Z \rightarrow \nu \bar{\nu})/dp^{Z}_{\rm{T}}-d\sigma(Z \rightarrow l^{+}l^{-})/dp^{Z}_{\rm{T}}}{d\sigma(Z \rightarrow \nu \bar{\nu})/dp^{Z}_{\rm{T}}+d\sigma(Z \rightarrow l^{+}l^{-})/dp^{Z}_{\rm{T}}}.
\end{equation}
Precise theoretical calculations of the $A_{\rm{inv}}$ can serve as a powerful probe for detection of excess as a result of new-physics signatures at an experiment, which would appear in final states of the $Z$ boson. The $A_{\rm{inv}}$ can also be sensitive to new-physics signatures in particular at large $p^{Z}_{\rm{T}}$ values, such as through 1 TeV region and beyond. Furthermore, the $A_{\rm{inv}}$ can be a good alternate of the $R_{\rm{inv}}$ in high-precision experimental studies by enabling cancellation of residual systematic uncertainties that are correlated in its ratio. The $A_{\rm{inv}}$ can also offer more sensitivity to Drell-Yan lepton-pair kinematics in comparison to the $R_{\rm{inv}}$ via distribution shapes which makes sense for new-resonance searches with high-$m_{ll}$ values in more constrained phase-space regions. After all to bear in mind that both the invisible probing observables $R_{\rm{inv}}$ and $A_{\rm{inv}}$ are simply referred to as the invisible probes in the context of this paper,   

To this end, the renormalization scale $\mu_{R}$ and the factorization scale $\mu_{F}$ which are required in calculations of hadronic cross sections for the invisible probes $R_{\rm{inv}}$ and $A_{\rm{inv}}$, are set equal to 

\begin{equation}
\label{eqn:5}   
\begin{array}{l}
\displaystyle
\mu_{\rm{R,F}}=\lambda_{\rm{R,F}}\it{\mu_{\rm{0}}}, \hspace{0.1cm} \rm{with} \hspace{0.1cm} \it{\mu_{\rm{0}}}=\it{m_{Z}}=\rm{91.1876} \hspace{0.1cm} \rm{GeV} \hspace{0.1cm} \rm{and} \hspace{0.1cm} 1/2\leq\lambda_{\rm{R}},\lambda_{\rm{F}}\leq 2,
\end{array}
\end{equation}   
where the $\lambda_{\rm{R}}=\lambda_{\rm{F}}=1$ condition refers to the physical $m_{Z}$ and thus the choice for the central values of the scales $\mu_{R}$ and $\mu_{F}$. This choice is used for the calculations which are acquired in the $Z$-boson resonance region $76<m_{ll}<106$ GeV, whereas central scale is set dynamically to invariant mass of lepton pair as $\mu_{\rm{0}}=m_{ll}$ for the calculations obtained in high-$m_{ll}$ region through the relevant parts of the paper. Theoretical uncertainties from missing higher-order QCD corrections beyond NNLO are estimated customarily using the 7-point variations of the scales as ($\lambda_{\rm{R}},\lambda_{\rm{F}}$)=(2,2), (2,1), (1,2), (1,1),(1,1/2), (1/2,1), (1/2,1/2). Largest uncertainty from up and down variations is taken as the theoretical scale uncertainty. Scale uncertainties are properly propagated to the differential distributions of the $R_{\rm{inv}}$ and $A_{\rm{inv}}$ by taking into account standard deviations. Numerator and denominator of the invisible probes are considered uncorrelated for the uncertainties assuming a conservative approach. 

\section{Phenomenological results}
\label{pheno}

\subsection{Comparison with the data}
\label{comp}
\begin{figure}
\center
\includegraphics[width=12cm]{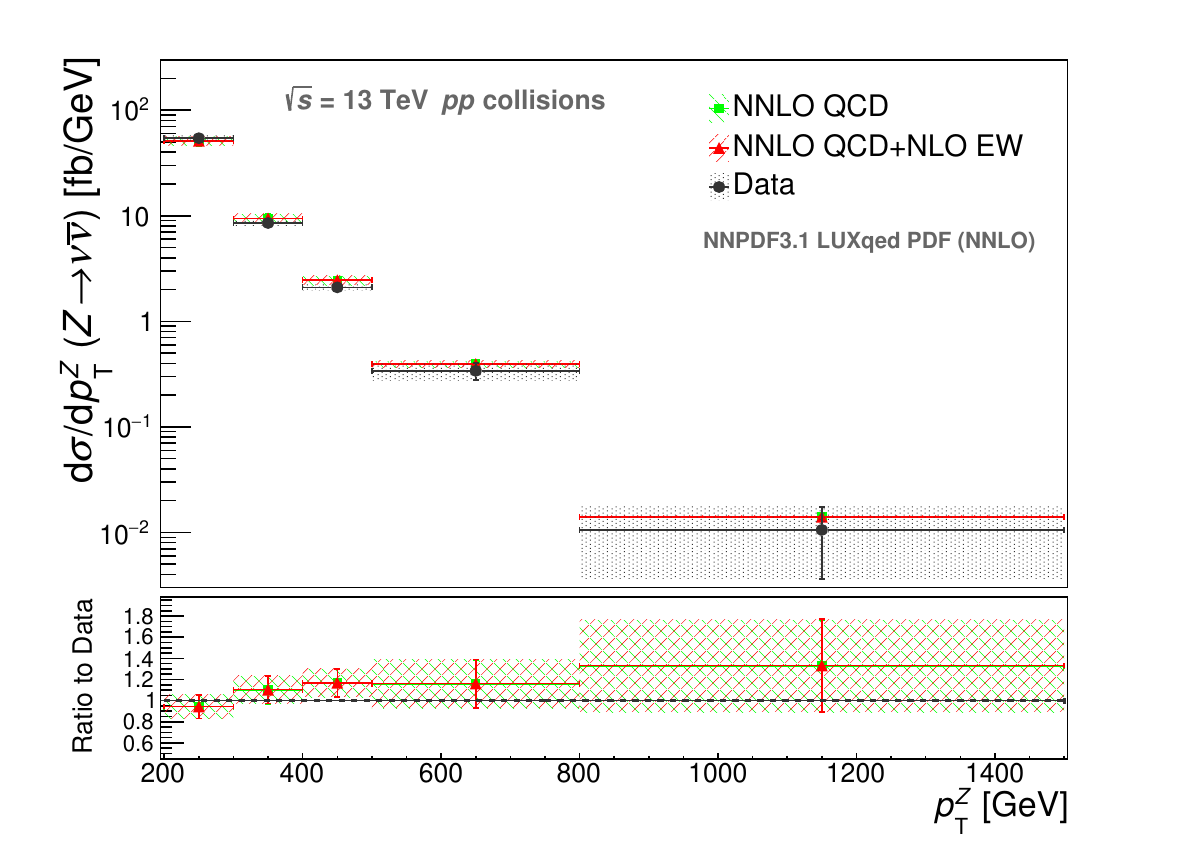}
\caption{The differential cross section distributions of the invisible $Z$-boson process by the NNLO QCD and the combined NNLO QCD+NLO EW predictions in comparison with the data at 13 TeV. The distributions are provided as a function of the $p^{Z}_{\rm{T}}$ in the range 200--1500 GeV. Theoretical uncertainties due to scale variations are included for the predictions and total experimental uncertainty is quoted for the data. In the lower panel, the ratios of the predictions to the data are shown.}
\label{fig:1}      
\end{figure}

Differential cross-section distributions of the invisible process $pp\rightarrow Z \rightarrow \nu \bar{\nu}$ are predicted at fixed-order NNLO QCD and combined NNLO QCD+NLO EW accuracies as a function of the $p^{Z}_{\rm{T}}$ at 13 TeV. The predicted distributions are obtained using the baseline selection for the invisible process with $p^{\rm{miss}}_{\rm{T}}>$ 200 GeV requirement as discussed in Sec.~\ref{meth}. The $p^{Z}_{\rm{T}}$ bin edges are chosen in the range 200--1500 GeV in line with the CMS measurement~\cite{CMS:2020hkd} to which the predicted distributions are compared. The predicted distributions in the last $p^{Z}_{\rm{T}}$ bin 800--1500 GeV is extrapolated from a wider bin range 500--1500 GeV to minimize numerical instabilities of the calculations. Scale uncertainties are quoted for the predictions in comparison with the data distribution including total experimental uncertainty. The predictions are compared with the CMS data for validation as shown in Figure~\ref{fig:1}. Difference between the NNLO QCD and the combined NNLO QCD+NLO EW predictions are observed to be small, amounting less than 0.5\%. The predictions are found to be in good agreement with the data throughout the entire $p^{Z}_{\rm{T}}$ range within quoted uncertainties. In the last two $p^{Z}_{\rm{T}}$ bins between 500--1500 GeV, the predictions are noticed to exhibit higher precision against larger experimental uncertainties. Predicted results tend to be centrally higher than the data in high-$p^{Z}_{\rm{T}}$ region, however they are precisely reliable for description of the data in this region of the invisible process.     

\begin{figure}
\center
\includegraphics[width=12cm]{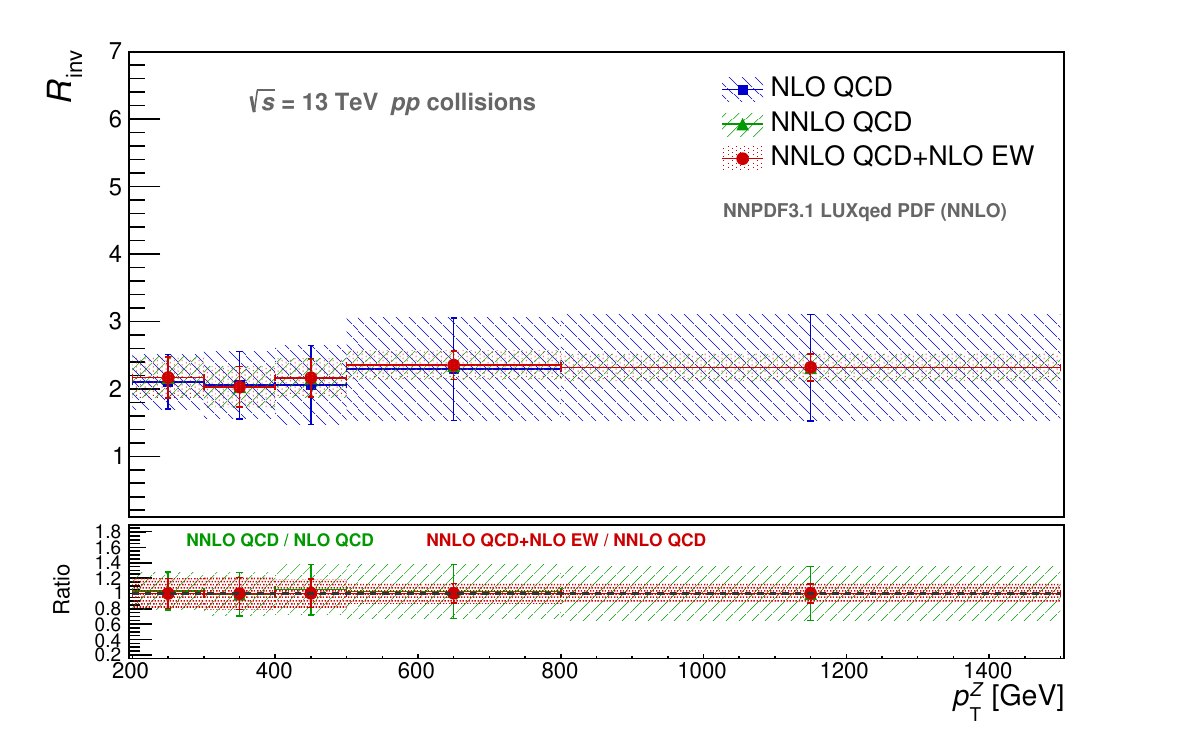}
\includegraphics[width=12cm]{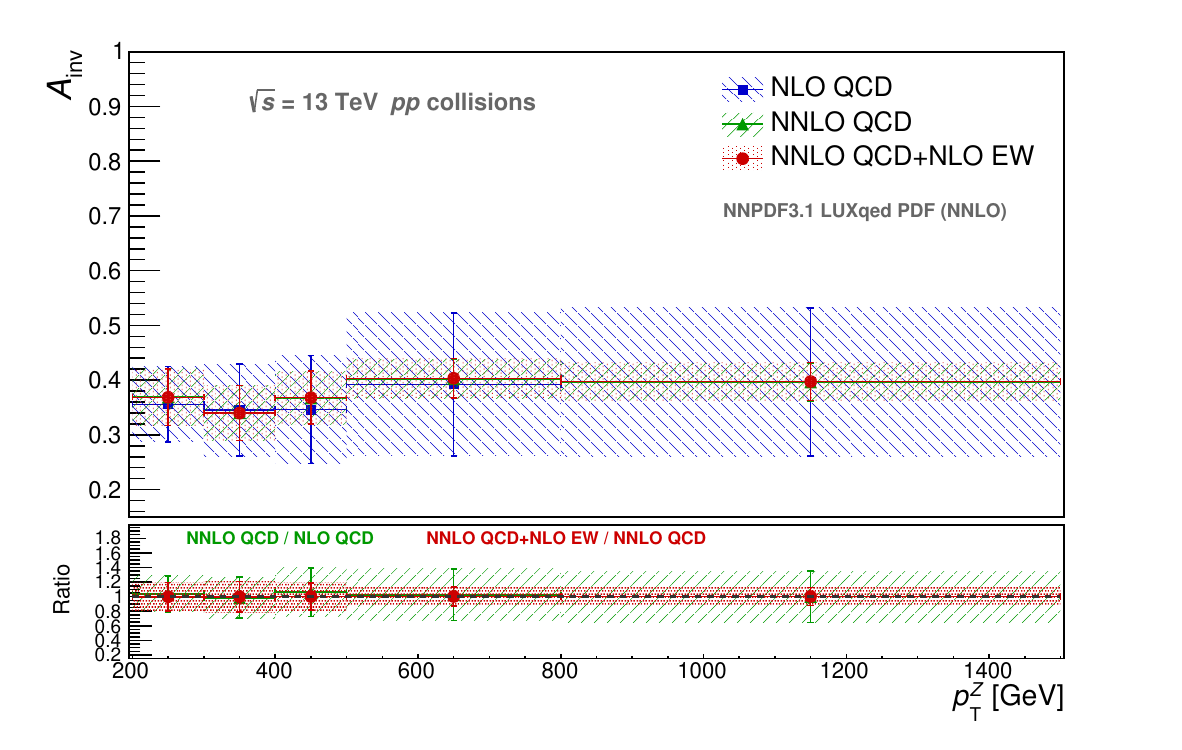}
\caption{The distributions of the invisible probes $R_{\rm{inv}}$ (top) and $A_{\rm{inv}}$ (bottom) by the (N)NLO QCD and NNLO QCD+NLO EW predictions at 13 TeV. The predicted distributions are presented in bins of the $p^{Z}_{\rm{T}}$ in the range 200--1500 GeV. Theoretical uncertainties due to scale variations are included for the predictions. The ratios of the predictions are provided in the lower panel.}
\label{fig:2}      
\end{figure}

Having presented justification of the predictions with the data, distributions for the invisible probes $R_{\rm{inv}}$ and $A_{\rm{inv}}$ as defined in Sec.~\ref{meth} are predicted at (N)NLO QCD and NNLO QCD+NLO EW at 13 TeV. The leptonic process $pp \rightarrow Z \rightarrow l^{+}l^{-}$ is considered to be in the $Z$-boson resonance region with the $m_{ll}$ window requirement $76<m_{ll}<106$ GeV and is required to have no $p^{l}_{\rm{T}}$ requirement based on the baseline selection as discussed in Sec.~\ref{meth}. The predicted distributions are obtained in bins of the $p^{Z}_{\rm{T}}$ in the range 200--1500 GeV and compared among different accuracies as shown in Figure~\ref{fig:2}. Predicted central values at NNLO QCD and NNLO QCD+NLO EW vary in the range 2.0--2.35 for the $R_{\rm{inv}}$, while they lie in the range 0.34--0.40 for the $A_{\rm{inv}}$. The predictions exhibit more flat behavior in the higher-$p^{Z}_{\rm{T}}$ range 500--1500 GeV. As a result, higher-$p^{Z}_{\rm{T}}$ regions of the invisible probes turn out to be a good kinematic probe for checking any deviation from the SM expectations by experimental results. Difference between NLO QCD and NNLO QCD predictions amounts to 4.8\% for the $R_{\rm{inv}}$, while it is up to 5.8\% for the $A_{\rm{inv}}$. Precision is significantly improved in going from NLO to NNLO QCD(+NLO EW) throughout the entire $p^{Z}_{\rm{T}}$ range. Difference between NNLO QCD and NNLO QCD+NLO EW are not sizable indicating cancellation of NLO EW effects inherently in ratios of the distributions. Numerical values by fixed-order NNLO QCD and the combined NNLO QCD+NLO EW predictions are also tabulated in Table~\ref{tab:1}.   

\begin{table*}
\caption{\label{tab:1}Numerical results of the invisible probes $R_{\rm{inv}}$ and $A_{\rm{inv}}$ in bins of the $p^{Z}_{\rm{T}}$ by the NNLO QCD and NNLO QCD+NLO EW predictions at 13 TeV. Theoretical uncertainties due to scale variations are included in the results.} 
\lineup
\resizebox{16.0cm}{!}{%
\begin{tabular}{@{}*{11}{c|cc|cc}}
\br
Probe & \multicolumn{2}{c|}{$R_{\rm{inv}}$} & \multicolumn{2}{c}{$A_{\rm{inv}}$}  \\
\mr
$p^{Z}_{\rm{T}}$ [GeV]  & $\rm{NNLO~QCD}$ & $\rm{NNLO~QCD\!+\!NLO~EW}$ & $\rm{NNLO~QCD}$ & $\rm{NNLO~QCD\!+\!NLO~EW}$    \\
\mr
200--300        &2.169$\pm$0.304	&	2.167$\pm$0.303	&	0.369$\pm$0.051       &	0.369$\pm$0.051		\cr
300--400        &2.031$\pm$0.301	&	2.028$\pm$0.300	&	0.340$\pm$0.050	  &	0.340$\pm$0.050		\cr
400--500 	      &2.161$\pm$0.285	&	2.165$\pm$0.285	&	0.367$\pm$0.048        &	0.368$\pm$0.048	        \cr
500--800	      &2.347$\pm$0.211      &	2.352$\pm$0.211	&	0.402$\pm$0.036  	  &	0.403$\pm$0.036	        \cr
800--1500	      &2.314$\pm$0.201	&	2.319$\pm$0.201	&	0.396$\pm$0.035  	  &	0.397$\pm$0.034		\cr
\br
\end{tabular}%
}
\end{table*}

\subsection{Impact of $p^{l}_{\rm{T}}$ threshold}
\label{impa}
\begin{figure} 
\center
\includegraphics[width=12cm]{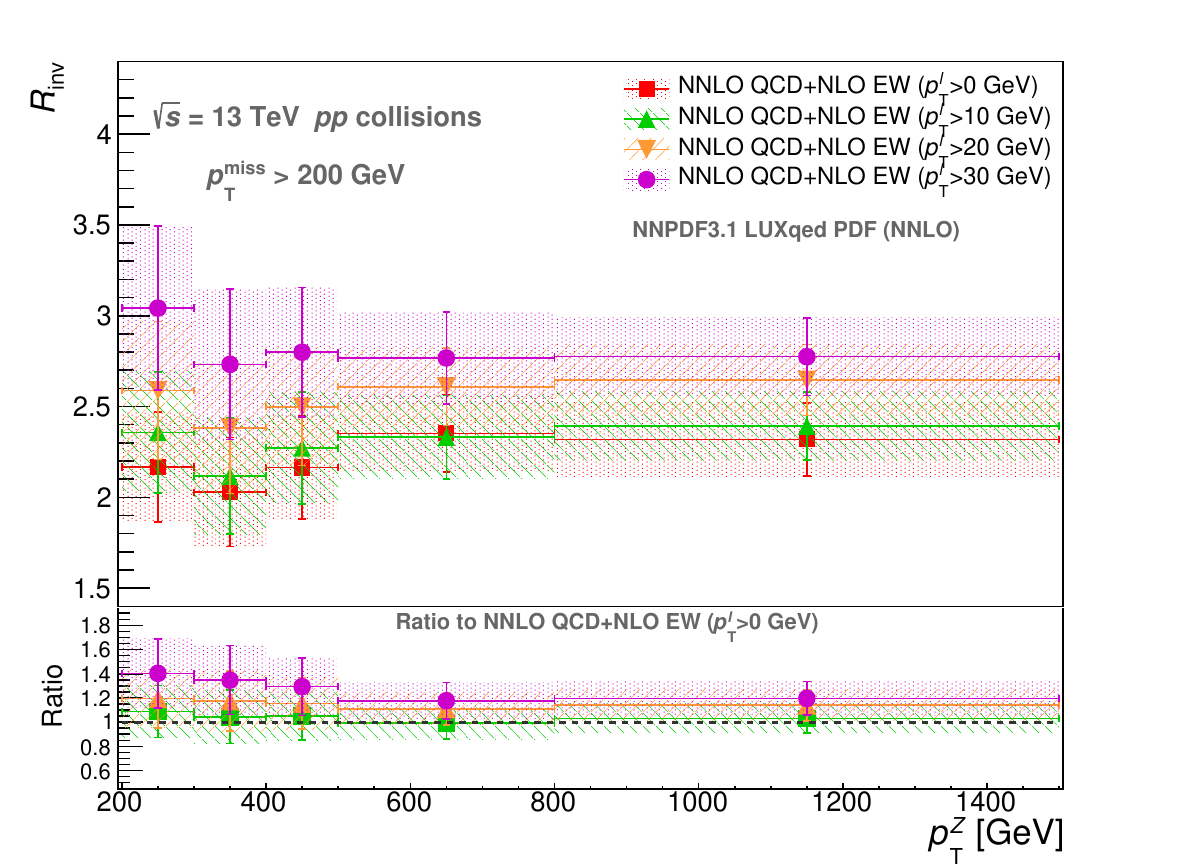}
\includegraphics[width=12cm]{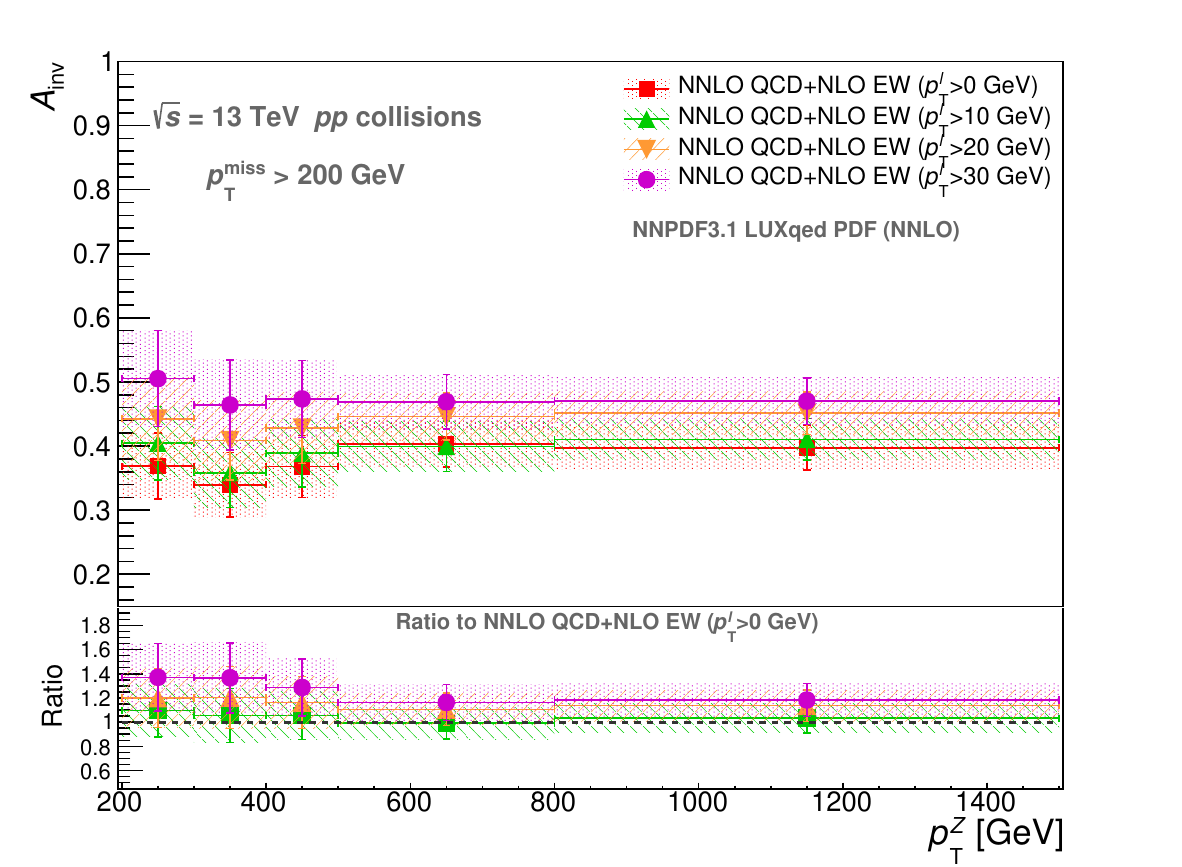}
\caption{The distributions of the invisible probes $R_{\rm{inv}}$ (top) and $A_{\rm{inv}}$ (bottom) by the NNLO QCD+NLO EW predictions at 13 TeV using $p^{l}_{\rm{T}}>$ 0, 10, 20, and 30 GeV threshold requirements. The predicted distributions are presented in bins of the $p^{Z}_{\rm{T}}$ in the range 200--1500 GeV. Theoretical uncertainties due to scale variations are included for the predictions. The ratios of the predictions with higher-$p^{l}_{\rm{T}}$ thresholds to the prediction with no $p^{l}_{\rm{T}}$ requirement are provided in the lower panel.}
\label{fig:3}      
\end{figure}

In the baseline selection of this paper, no $p^{l}_{\rm{T}}$-threshold requirement (i.e., $p^{l}_{\rm{T}}>$ 0) is imposed for the leptonic $Z$-boson process. Nevertheless, any threshold requirements for the $p^{l}_{\rm{T}}$ will result in a more constrained phase space for the leptonic process and impact predictions of the $R_{\rm{inv}}$ and $A_{\rm{inv}}$ for controlling the invisible $Z$-boson process and probing new-physics scenarios. Therefore, it is of substantial interest to predict differential distributions of the probes $R_{\rm{inv}}$ and $A_{\rm{inv}}$ in the presence of successive threshold requirements $p^{l}_{\rm{T}}>$ 10, 20, and 30 GeV. By this token, predicted distributions with these $p^{l}_{\rm{T}}$ requirements at 13 TeV are acquired in bins of the $p^{Z}_{\rm{T}}$ based on the combined NNLO QCD+NLO EW accuracy. NNLO QCD+NLO EW predictions for the $R_{\rm{inv}}$ and $A_{\rm{inv}}$ are obtained in the $Z$-boson resonance $76<m_{ll}<106$ GeV and are compared in Figure~\ref{fig:3} for different threshold requirements $p^{l}_{\rm{T}}>$ 0, 10, 20, and 30 GeV. The requirement $p^{\rm{miss}}_{\rm{T}}>$ 200 GeV is preserved in the selection of the invisible process in the predictions. The predictions are clearly observed to exhibit higher $R_{\rm{inv}}$ and $A_{\rm{inv}}$ distributions in going from the $p^{l}_{\rm{T}}>$ 0 to $p^{l}_{\rm{T}}>$ 30 GeV requirement. The impact of higher-$p^{l}_{\rm{T}}$ threshold on the invisible probes is more pronounced in the low-$p^{Z}_{\rm{T}}$ range 200--500 GeV, which is up to $\sim$40\% for the central points. On the other hand, the impact of higher-$p^{l}_{\rm{T}}$-threshold requirements is up to $\sim$20\% regarding the central values of the invisible probes in the high--$p^{Z}_{\rm{T}}$ range 500--1500 GeV. In this high--$p^{Z}_{\rm{T}}$ range, the $R_{\rm{inv}}$ and $A_{\rm{inv}}$ distributions become more flat in the predictions with the $p^{l}_{\rm{T}}>$ 30 GeV threshold requirement. Both of the invisible probes with varying low-$p^{l}_{\rm{T}}$ thresholds can maintain their sensitivities to even small deviation of experimental data from the SM predictions which would mimic new-physics phenomena in the high--$p^{Z}_{\rm{T}}$ region. Particularly any considerable excess over the predicted $A_{\rm{inv}}$ distribution with higher-$p^{l}_{\rm{T}}$ thresholds by experimental data in the high--$p^{Z}_{\rm{T}}$ region can be translated into searches of signs for dark-matter candidates. Finally, it has been noted that the prediction with the $p^{l}_{\rm{T}}>$ 30 GeV has the largest $R_{\rm{inv}}$ and $A_{\rm{inv}}$ values in the first $p^{Z}_{\rm{T}}$ bin 200--300 GeV in contrast to the predictions with lower-$p^{l}_{\rm{T}}$ thresholds, as they have their largest values towards the higher-$p^{Z}_{\rm{T}}$ region of phase space. 
 
\subsection{Invisible distributions in high-$m_{ll}$ region}
\label{highmass}
Pursuit of both resonant and nonresonant signatures in high-invariant mass region of Drell-Yan lepton-pair production constitutes an integral part of new-physics searches at the LHC~\cite{CMS:2021ctt,ATLAS:2019erb}. High-$m_{ll}$ region of lepton-pair final states beyond the $Z$-boson resonance, i.e., $76<m_{ll}<106$ GeV as used in this paper, can be probed by using the invisible $Z$-boson process which will contribute to extension of new-physics searches. Differential distributions of the invisible probes $R_{\rm{inv}}$ and $A_{\rm{inv}}$ as introduced before, are proposed for exploring high-$m_{ll}$ region of lepton pair emitted in final states. For this purpose, higher-$m_{ll}$ regions $106<m_{ll}<350$ GeV and $350<m_{ll}<1000$ GeV are required in the predictions, apart from the baseline selection with the $76<m_{ll}<106$ GeV requirement, in the selection of leptonic decay of the $Z$ boson corresponding to lepton-pair process in final states. The requirement $p^{l}_{\rm{T}}>$ 0 (meaning to no $p^{l}_{\rm{T}}$ requirement) is preserved in the selection of the leptonic process as well the requirement $p^{\rm{miss}}_{\rm{T}}>$ 200 GeV in the selection of the invisible process. The predicted $R_{\rm{inv}}$ and $A_{\rm{inv}}$ distributions are compared among the $Z$-boson resonance region and the higher-$m_{ll}$ regions based on the combined calculations of the NNLO QCD and NLO EW corrections of the perturbation theories. 

\begin{figure} 
\center
\includegraphics[width=12cm]{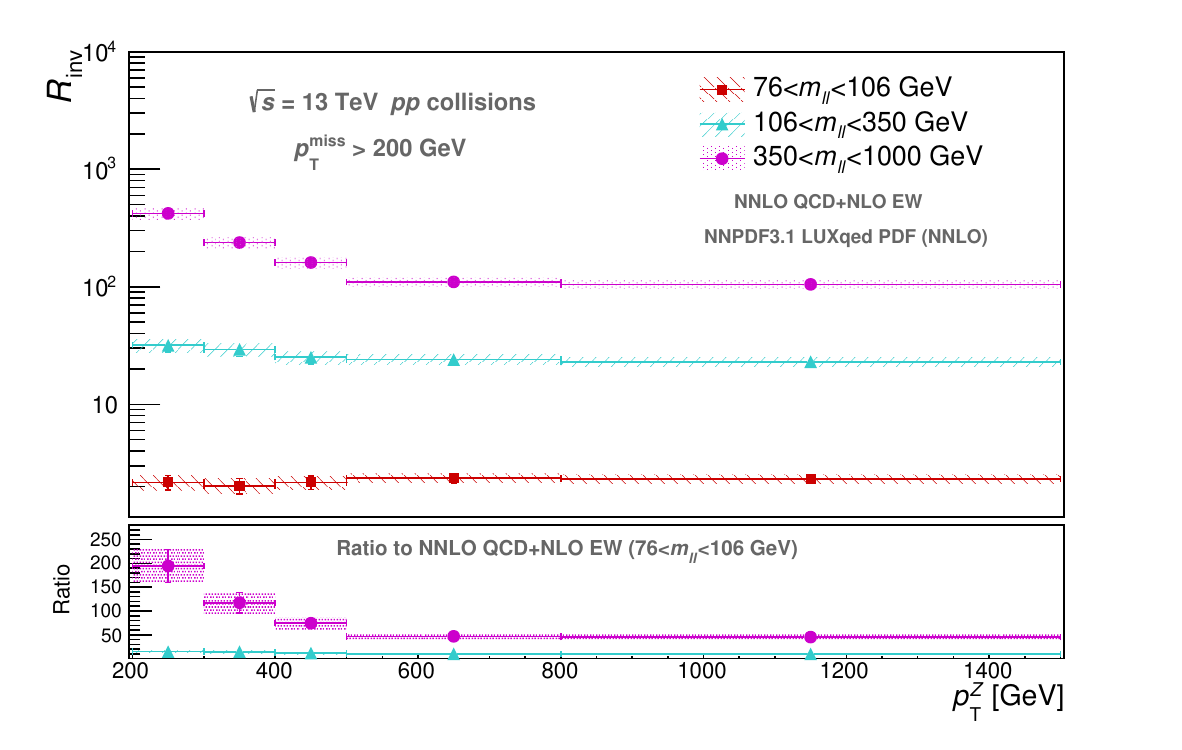}
\includegraphics[width=12cm]{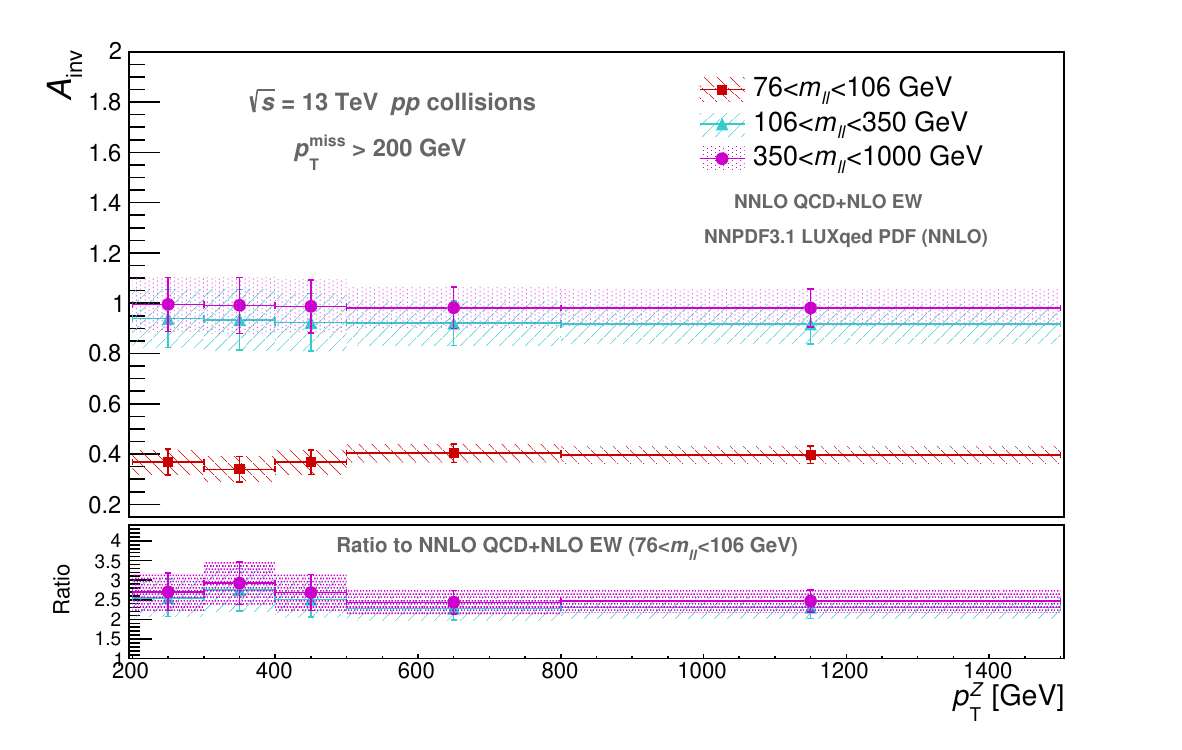}
\caption{The distributions of the invisible probes $R_{\rm{inv}}$ (top) and $A_{\rm{inv}}$ (bottom) by the NNLO QCD+NLO EW predictions at 13 TeV using $76<m_{ll}<106$ GeV, $106<m_{ll}<350$ GeV, and $350<m_{ll}<1000$ GeV requirements. The predicted distributions are presented in bins of the $p^{Z}_{\rm{T}}$ in the range 200--1500 GeV. Theoretical uncertainties due to scale variations are included for the predictions. The ratios of the predictions of higher-$m_{ll}$ regions to the prediction of the $Z$-boson resonance region $76<m_{ll}<106$ GeV are provided in the lower panel.}
\label{fig:4}      
\end{figure}

The predicted distributions with different $m_{ll}$ region requirements are compared in bins of the $p^{Z}_{\rm{T}}$ as shown in Figure~\ref{fig:4}. The $R_{\rm{inv}}$ distribution increases significantly from the resonance region onwards the highest-$m_{ll}$ region 350--1000 GeV. The $R_{\rm{inv}}$ distribution is consistently more flat in the resonance region for the entire $p^{Z}_{\rm{T}}$ range, whereas it starts to become flat in the high-$m_{ll}$ regions 106--1000 GeV for only the $p^{Z}_{\rm{T}}$ range 500--1500 GeV. The $R_{\rm{inv}}$ distribution is the highest in the first $p^{Z}_{\rm{T}}$ bin of the highest-$m_{ll}$ region 350--1000 GeV, which is observed to be $\sim$421.0. In the highest-$m_{ll}$ region, the distribution decreases gradually towards higher-$p^{Z}_{\rm{T}}$ ranges and becomes almost flat between the values 104.0--110.0. Moreover, the $A_{\rm{inv}}$ distribution is maximized in going from the resonance region to the highest-$m_{ll}$ region, where it amounts up to $\sim$0.99 in the lowest three $p^{Z}_{\rm{T}}$ bins and up to $\sim$0.98 in the $p^{Z}_{\rm{T}}$ range 500--1500 GeV. The $A_{\rm{inv}}$ distribution is lowest around $\sim$0.34 in the resonance region as elaborated also in the previous subsections. The invisible probe $A_{\rm{inv}}$ is anticipated to increase more to be $\sim$1.0 for an $m_{ll}$ region greater than 1 TeV. In precision measurements, any significant deviation which will violate this almost asymmetrical behaviour of the $A_{\rm{inv}}$ (i.e., $A_{\rm{inv}}\sim$1.0 with an experimental uncertainty less than percent level) in a region of phase space of lepton-pair production with $m_{ll}>$1 TeV and primarily towards the lower $p^{Z}_{\rm{T}}$ region of 200--400 GeV would stand for considerable hints for new physics. Apart from this discussion and in general, both the $R_{\rm{inv}}$ and $A_{\rm{inv}}$ distributions in high-$m_{ll}$ region 106--1000 GeV can be considered as good probes of Drell-Yan lepton-pair final states. To this end, numerical values from the predicted distributions of the invisible probes are provided in Table~\ref{tab:2} for detailed comparisons among the resonance region and high-$m_{ll}$ regions.          

\begin{table*}
\caption{\label{tab:2}Numerical results of the invisible probes $R_{\rm{inv}}$ and $A_{\rm{inv}}$ in bins of the $p^{Z}_{\rm{T}}$ at NNLO QCD+NLO EW accuracy at 13 TeV. The probes are predicted in the high-$m_{ll}$ regions $106<m_{ll}<350$ GeV and $350<m_{ll}<1000$ GeV in comparison with the $Z$-boson resonance region $76<m_{ll}<106$ GeV. Theoretical uncertainties due to scale variations are included in the results.} 
\lineup
\resizebox{16.0cm}{!}{%
\begin{tabular}{@{}*{11}{c|ccc|ccc}}
\br
Probe & \multicolumn{3}{c|}{$R_{\rm{inv}}$} & \multicolumn{3}{c}{$A_{\rm{inv}}$}  \\
\mr
$p^{Z}_{\rm{T}}$ [GeV]  & 76--106 GeV & 106--350 GeV & 350--1000 GeV & 76--106 GeV & 106--350 GeV & 350--1000 GeV   \\
\mr
200--300        &2.167$\pm$0.303	&	31.787$\pm$3.921	&	421.378$\pm$45.529     &	  0.369$\pm$0.051	 &   0.940$\pm$0.116     &	  0.995$\pm$0.107  	  \cr
300--400        &2.028$\pm$0.300	&	29.343$\pm$3.805	&	237.917$\pm$26.682     &   0.340$\pm$0.050	 &   0.934$\pm$0.121     &	  0.992$\pm$0.111	  \cr
400--500 	      &2.165$\pm$0.285	&	25.143$\pm$3.119 	&	160.875$\pm$17.164     &	  0.368$\pm$0.048	 &   0.924$\pm$0.115     &	  0.988$\pm$0.105      \cr
500--800	      &2.352$\pm$0.211      &	24.072$\pm$2.296	&	110.030$\pm$9.170       &	  0.403$\pm$0.036	 &   0.920$\pm$0.088     &	  0.982$\pm$0.082      \cr
800--1500	      &2.319$\pm$0.201	&	23.014$\pm$1.967	&	104.795$\pm$8.019       &	  0.397$\pm$0.034	 &   0.917$\pm$0.078     &	  0.981$\pm$0.075	   \cr
\br
\end{tabular}%
}
\end{table*}

\subsection{High-$p^{\rm{miss}}_{\rm{T}}$ invisible probes for high-$m_{ll}$ region}
\label{highptmiss}
In continuation of tests of high-$m_{ll}$ phase space region of Drell-Yan lepton-pair process, it is of much interest to accommodate more probing power by the invisible $Z$-boson process with higher-$p^{\rm{miss}}_{\rm{T}}$ requirements. The neutrino pair emitted with higher-$p^{\rm{miss}}_{\rm{T}}$ result in higher missing transverse momentum resolution and pay the way for better identification of the invisible process at an experiment. Therefore, the invisible process at large $p^{\rm{miss}}_{\rm{T}}$ with better experimental handles can be benefited for probing lepton-pair process in high-$m_{ll}$ region. Accordingly the predictions for the invisible process are required to have $p^{\rm{miss}}_{\rm{T}}>$ 300 and 400 GeV in addition to the prediction with the baseline requirement $p^{\rm{miss}}_{\rm{T}}>$ 200 GeV. First of all, numerical values of the $R_{\rm{inv}}$ and $A_{\rm{inv}}$ predictions are given in Table~\ref{tab:3} for the $Z$-boson resonance region $76<m_{ll}<106$ GeV using the successive requirements $p^{\rm{miss}}_{\rm{T}}>$200, 300, and 400 GeV. The $R_{\rm{inv}}$ varies generally by either increasing or decreasing in going from $p^{\rm{miss}}_{\rm{T}}>$200 GeV to higher-$p^{\rm{miss}}_{\rm{T}}$ requirements by up to $\sim$1.2\%. The impact of $p^{\rm{miss}}_{\rm{T}}>$400 GeV requirement relative to the $p^{\rm{miss}}_{\rm{T}}>$200 GeV requirement is only $\sim$0.2\% in the $p^{Z}_{\rm{T}}$ range 500--1500 GeV, corresponding to a phase space region with more flat $R_{\rm{inv}}$ distribution. Similarly in the $Z$-boson resonance, the $p^{\rm{miss}}_{\rm{T}}>$400 GeV requirement affects the $A_{\rm{inv}}$ by up to $\sim$0.2\% relative to the $p^{\rm{miss}}_{\rm{T}}>$200 GeV requirement in the high-$p^{Z}_{\rm{T}}$ range 500--1500 GeV. Therefore, the invisible probes are found to be more sensitive to increasing $p^{\rm{miss}}_{\rm{T}}$ values in the low-$p^{Z}_{\rm{T}}$ range 200--500 GeV in comparison to the region with $p^{Z}_{\rm{T}}>$ 500 GeV.   

\begin{table*}
\caption{\label{tab:3}Numerical results of the probes $R_{\rm{inv}}$ and $A_{\rm{inv}}$ in bins of the $p^{Z}_{\rm{T}}$ at NNLO QCD+NLO EW accuracy at 13 TeV. The predicted results are obtained in the $Z$-boson resonance region $76<m_{ll}<106$ GeV with the successive requirements $p^{\rm{miss}}_{\rm{T}}>$200, 300, and 400 GeV imposed for the invisible process. Theoretical uncertainties due to scale variations are associated with the central results.} 
\lineup
\resizebox{16.0cm}{!}{%
\begin{tabular}{@{}*{11}{c|ccc|ccc}}
\br
Probe & \multicolumn{3}{c|}{$R_{\rm{inv}}$}  & \multicolumn{3}{c}{$A_{\rm{inv}}$}\\
\mr
$p^{Z}_{\rm{T}}$ [GeV]  & $p^{\rm{miss}}_{\rm{T}}>$ 200 GeV & $>$ 300 GeV & $>$ 400 GeV & $p^{\rm{miss}}_{\rm{T}}>$ 200 GeV & $>$ 300 GeV & $>$ 400 GeV   \\
\mr
200--300        &2.167$\pm$0.302 	&	--	                                 &	--                                    &	  0.369$\pm$0.051	 &   --                                 &	 -- 	                             \cr
300--400        &2.028$\pm$0.300 	&	2.015$\pm$0.294	        &	--                                    &   0.340$\pm$0.050	 &   0.337$\pm$0.164       &	  --	                              \cr
400--500 	      &2.164$\pm$0.285	&	2.135$\pm$0.276 	        &	2.141$\pm$0.273           &	  0.368$\pm$0.048	 &   0.362$\pm$0.052       &	  0.363$\pm$0.046       \cr
500--800	      &2.352$\pm$0.211      &	2.381$\pm$0.217   	        &	2.347$\pm$0.209           &	  0.403$\pm$0.036	 &   0.408$\pm$0.024       &	  0.402$\pm$0.036       \cr
800--1500	      &2.319$\pm$0.202  	&	2.345$\pm$0.207   	        &	2.317$\pm$0.200           &	  0.397$\pm$0.035     &   0.402$\pm$0.024       &	  0.397$\pm$0.034	    \cr
\br
\end{tabular}%
}
\end{table*}

Next, two high-$m_{ll}$ regions are required for the leptonic process as $350<m_{ll}<1000$ GeV and $106~\rm{GeV}<m_{ll}<\infty$ to assess impact of higher $p^{\rm{miss}}_{\rm{T}}$ requirements. Phase spaces $\wp(i,j)$ of the invisible probes $R_{\rm{inv}}$ and $A_{\rm{inv}}$ are then formed suitably by combining the $p^{\rm{miss}}_{\rm{T}}$ and $m_{ll}$ requirements based on the following representation  

\begin{equation}
\label{eqn:6}   
\begin{array}{l}
\displaystyle
\wp(i,j)=\wp_{i}(p^{\rm{miss}}_{\rm{T}})+\wp_{j}(m_{ll})  \\
\rm{with}~\wp_{i=1,2,3}(\it{p}^{\rm{miss}}_{\rm{T}})=\wp(p^{\rm{miss}}_{\rm{T}}>\rm{200},\rm{300},\rm{400}~\rm{GeV})  \\
\rm{and}~\wp_{j=1,2}(\it{m_{ll}})=\wp(\rm{350}<\it{m_{ll}}<\rm{1000}~GeV,\rm{106}~GeV<\it{m_{ll}}<\infty), 
\end{array}
\end{equation} 
such as $\wp(1,1)$ specifies a phase space with $p^{\rm{miss}}_{\rm{T}}>$ 200 GeV and $350<m_{ll}<1000$ GeV selection and so on. The predicted numerical values at NNLO QCD+NLO EW accuracy for the $R_{\rm{inv}}$ and $A_{\rm{inv}}$ are reported in the combined phase spaces in Tables~\ref{tab:4} and~\ref{tab:5}, respectively. Numerical values are not evaluated for $\wp(2,1)$ and $\wp(2,2)$ in the $p^{Z}_{\rm{T}}$ range 200--300 GeV owing to $p^{\rm{miss}}_{\rm{T}}>$ 300 GeV requirement of the invisible process, similarly for $\wp(3,1)$ and $\wp(3,2)$ in the $p^{Z}_{\rm{T}}$ range 200--400 GeV due to $p^{\rm{miss}}_{\rm{T}}>$ 400 GeV requirement. In going from $p^{\rm{miss}}_{\rm{T}}>$ 200 GeV to 300 and 400 GeV, the probe $R_{\rm{inv}}$ varies up to $\sim$1.2\%. The predicted results show that difference in $R_{\rm{inv}}$ is reduced to be very small $\sim$0.2\% at increasingly large $p^{\rm{miss}}_{\rm{T}}$ values in the region $p^{Z}_{\rm{T}}>$500 GeV. As a consequence, the $R_{\rm{inv}}$ in a phase space region with sufficiently large $p^{\rm{miss}}_{\rm{T}}$ value can sustain to be a good probe for lepton-pair process in the high-$m_{ll}$ region beyond the $Z$-boson resonance. The $R_{\rm{inv}}$ values increase significantly towards higher-$m_{ll}$ region relative to the $Z$-boson resonance, assuring lepton-pair process to be potentially much more interesting for new-physics searches at large $m_{ll}$ values. As discussed before in Sec.~\ref{highmass} the $A_{\rm{inv}}$ increases directly with requirement of higher-$m_{ll}$ region of lepton-pair process. Nevertheless, the $A_{\rm{inv}}$ values are predicted to be almost unchanged with higher-$p^{\rm{miss}}_{\rm{T}}$ requirements relative to the $p^{\rm{miss}}_{\rm{T}}>$ 200 GeV requirement. No matter how harder neutrino pair is emitted with large $p^{\rm{miss}}_{\rm{T}}$ values, the $A_{\rm{inv}}$ tends to be unaffected in the high-$m_{ll}$ region in contrast to $Z$-boson resonance to be more powerful probe and provide potentially more reliable control. Therefore, any deviation from the $A_{\rm{inv}}$ probe with increasing $p^{\rm{miss}}_{\rm{T}}$ values in the high-$m_{ll}$ region can be considered as a reliable signs in searches of new-physics scenarios through both invisible decay product or lepton-pair final-states.         

\begin{table*}
\caption{\label{tab:4}Numerical results of the $R_{\rm{inv}}$ in bins of the $p^{Z}_{\rm{T}}$ at NNLO QCD+NLO EW accuracy at 13 TeV. The predicted results are obtained in 6 different phase space regions $\wp(i,j)$ corresponding to combinations of $p^{\rm{miss}}_{\rm{T}}$ and $m_{ll}$ requirements as defined in Equation~\ref{eqn:6}. Theoretical uncertainties due to scale variations are associated with the central results.} 
\lineup
\resizebox{16.0cm}{!}{%
\begin{tabular}{@{}*{11}{c|ccc|ccc}}
\br
Probe & \multicolumn{6}{c}{$R_{\rm{inv}}$}  \\
\mr
$p^{Z}_{\rm{T}}$ [GeV]  & $\wp(1,1)$ & $\wp(2,1)$ & $\wp(3,1)$ & $\wp(1,2)$ & $\wp(2,2)$ & $\wp(3,2)$   \\
\mr
200--300        &421.378$\pm$45.529 	&	--	                                 &	--                                    &	  29.407$\pm$3.626	 &   --                                &	 -- 	                             \cr
300--400        &237.917$\pm$26.682 	&	236.379$\pm$25.833	&	--                                    &   26.136$\pm$3.334	 &   25.967$\pm$3.248     &	  --	                              \cr
400--500 	      &160.875$\pm$17.164	&	158.708$\pm$16.478 	&	159.096$\pm$16.173     &	  22.113$\pm$2.714	 &   21.815$\pm$2.623     &	  21.868$\pm$2.589      \cr
500--800	      &110.029$\pm$9.170          &	111.382$\pm$9.465   	&	109.815$\pm$9.038       &	  18.755$\pm$1.844	 &   18.986$\pm$1.893     &	  18.719$\pm$1.824      \cr
800--1500	      &104.795$\pm$8.019  	&	105.997$\pm$8.276   	&	104.711$\pm$7.919       &	  18.718$\pm$1.686	 &   18.933$\pm$1.731     &	  18.703$\pm$1.671	    \cr
\br
\end{tabular}%
}
\end{table*}

\begin{table*}
\caption{\label{tab:5}Numerical results of the $A_{\rm{inv}}$ in bins of the $p^{Z}_{\rm{T}}$ at NNLO QCD+NLO EW accuracy at 13 TeV. The predicted results are obtained in 6 different phase space regions $\wp(i,j)$ corresponding to combinations of $p^{\rm{miss}}_{\rm{T}}$ and $m_{ll}$ requirements as defined in Equation~\ref{eqn:6}. Theoretical uncertainties due to scale variations are associated with the central results.} 
\lineup
\resizebox{16.0cm}{!}{%
\begin{tabular}{@{}*{11}{c|ccc|ccc}}
\br
Probe & \multicolumn{6}{c}{$A_{\rm{inv}}$}  \\
\mr
$p^{Z}_{\rm{T}}$ [GeV]  & $\wp(1,1)$ & $\wp(2,1)$ & $\wp(3,1)$ & $\wp(1,2)$ & $\wp(2,2)$ & $\wp(3,2)$   \\
\mr
200--300        &0.995$\pm$0.108	&	--	                         &	--                                &	  0.934$\pm$0.115	 &   --                              &	  --  	                           \cr
300--400        &0.992$\pm$0.111	&	0.992$\pm$0.108	&	--                                &       0.926$\pm$0.118	 &   0.926$\pm$0.116     &	  --	                           \cr
400--500 	      &0.988$\pm$0.105	&	0.987$\pm$0.103 	&	0.988$\pm$0.100       &	  0.913$\pm$0.112	 &   0.912$\pm$0.110     &	  0.913$\pm$0.108      \cr
500--800	      &0.982$\pm$0.082      &	0.982$\pm$0.083	&	0.982$\pm$0.081       &	  0.899$\pm$0.088	 &   0.900$\pm$0.090     &	  0.899$\pm$0.088      \cr
800--1500	      &0.981$\pm$0.075	&	0.981$\pm$0.077	&	0.981$\pm$0.074       &	  0.899$\pm$0.081	 &   0.900$\pm$0.082     &	  0.898$\pm$0.080	   \cr
\br
\end{tabular}%
}
\end{table*}
		
\section{Conclusions}
\label{conclusion}
This paper presents genuine higher-order predictions accurate up to NNLO QCD+NLO EW for the invisible $Z$-boson decay process $Z \rightarrow \nu \bar{\nu}$ relative to the leptonic $Z$-boson decay process $Z \rightarrow l^{+}l^{-}$ in pp collisions at 13 TeV. The predictions are validated with the available LHC data for the differential cross section distribution of the invisible process as a function of the $p^{Z}_{\rm{T}}$. For the first time, differential cross-section ratio $R_{\rm{inv}}$ and asymmetry $A_{\rm{inv}}$ observables are introduced as a function of the $p^{Z}_{\rm{T}}$ for the invisible process relative to the leptonic process. The $R_{\rm{inv}}$ and $A_{\rm{inv}}$ observables are referred to as the invisible probes for their powerful aspects in controlling the invisible decay from the well-measured leptonic process and probing indirectly signs for new-physics searches beyond the SM. The invisible probes $R_{\rm{inv}}$ and $A_{\rm{inv}}$ are observed to have generally flat distributions in the high-$p^{Z}_{\rm{T}}$ region 500--1500 GeV, where they could provide valuable inputs in searches of deviation from the SM predictions with higher-precision measurements. The invisible probes are assessed to depend notably on the low-$p^{l}_{\rm{T}}$ thresholds between 0--30 GeV in the low-$p^{Z}_{\rm{T}}$ region 200--500 GeV, whereas they are observed to be good probes regardless of $p^{l}_{\rm{T}}$ threshold with more flat distributions in the high-$p^{Z}_{\rm{T}}$ region 500--1500 GeV. The predictions are reported in high-$m_{ll}$ region 106--1000 GeV of Drell-Yan lepton-pair process beyond the $Z$-boson mass resonance 76--106 GeV, where the $R_{\rm{inv}}$ and $A_{\rm{inv}}$ are proposed as important probes for searches of new-physics models in lepton-pair final states. The invisible probes are found to increase significantly in the high-$m_{ll}$ region 106--1000 GeV, where the $A_{\rm{inv}}$ is observed to increase towards $\sim$1.0. The $A_{\rm{inv}}$ is therefore shown to be a powerful probe, meaning that any violation of asymmetrical behavior of the $A_{\rm{inv}}$ through a region $m_{ll}>$ 1 TeV would be a considerable evidence for new-physics particles emitted in final states of the invisible process or the lepton-pair process. Moreover, the invisible probes are shown to vary at most $\sim$1.0\% level in the $Z$-boson resonance 76--106 GeV with increasing $p^{\rm{miss}}_{\rm{T}}$ thresholds between 200--400 GeV. The $R_{\rm{inv}}$ is impacted by increasing $p^{\rm{miss}}_{\rm{T}}$ thresholds between 200--400 GeV up to $\sim$1.0\% level in the high-$m_{ll}$ regions 350--1000 GeV and 106--$\infty$, whereas the impact on the $A_{\rm{inv}}$ is assessed to be negligible with higher-$p^{\rm{miss}}_{\rm{T}}$ requirements. The $A_{\rm{inv}}$ could potentially be less sensitive to mis-measurements of the invisible final states at an experiment which impacts $p^{\rm{miss}}_{\rm{T}}$. By this token, the $A_{\rm{inv}}$ can be considered to be a reliable probe for indirect searches in high-$m_{ll}$ region of lepton-pair final states. Conclusively the invisible probes $R_{\rm{inv}}$ and $A_{\rm{inv}}$ are not only important for controlling the invisible decay of the $Z$ boson, but they are also proposed to be important for probing deviation from the SM predictions or even putting some constraints in new-physics searches.

\ack{}
The numerical calculations reported in this paper were fully performed by using High Performance and Grid Computing Center (TRUBA resources) at TUBITAK ULAKBIM.  

\section*{Conflict of interest}
The author declares no conflict of interest throughout the entire paper.


\section*{References}
\begin{thebibliography}{35}


\bibitem{ATLAS:2016fij}
G. Aad et al., Measurement of $W^{\pm}$ and $Z$-boson production cross sections in $pp$ collisions at $\sqrt{s}=13$ TeV with the ATLAS detector, Phys. Lett. B, 759, 601--621 (2016)
\bibitem{ATLAS:2019zci}
G. Aad et al., Measurement of the transverse momentum distribution of Drell\textendash{}Yan lepton pairs in proton\textendash{}proton collisions at $\sqrt{s}=13$ TeV with the ATLAS detector, Eur. Phys. J. C, 80, 616 (2020)
\bibitem{CMS:2018mdl}
A. M. Sirunyan et al., Measurement of the differential Drell-Yan cross section in proton-proton collisions at $\sqrt{s}=13$ TeV, J. High Energy Phys., 12, 059 (2019)
\bibitem{CMS:2019raw}
A. M. Sirunyan et al., Measurements of differential Z boson production cross sections in proton-proton collisions at $\sqrt{s} $ = 13 TeV, J. High Energy Phys., 12, 061 (2019)
\bibitem{LHCb:2021huf}
R. Aaij et al., Precision measurement of forward $Z$ boson production in proton-proton collisions at $\sqrt{s} = 13$ TeV, J. High Energy Phys., 07, 026 (2022)

\bibitem{Kuhn:2005az}
J. H. Kuhn, A. Kulesza, S. Pozzorini, and M. Schulze, One-loop weak corrections to hadronic production of Z bosons at large transverse momenta, Nucl. Phys. B, 727, 368--394 (2005) 
\bibitem{Denner:2011vu}
A. Denner, S. Dittmaier, T. Kasprzik, and A. Muck, Electroweak corrections to dilepton + jet production at hadron colliders, JHEP, 06, 069 (2011) 
\bibitem{Dittmaier:2014qza}
S. Dittmaier, A. Huss, and C. Schwinn, Mixed QCD-electroweak $\mathcal{O}(\alpha_s\alpha)$ corrections to Drell-Yan processes in the resonance region: pole approximation and non-factorizable corrections, Nucl. Phys. B, 885, 318--372 (2014)
\bibitem{Lindert:2017olm}
J. M. Lindert et al., Precise predictions for $V+$ jets dark matter backgrounds, Eur. Phys. J. C, 77, 829 (2017) 

\bibitem{Hamberg:1990np}
R. Hamberg, W. L. van Neerven, and T. Matsuura, A complete calculation of the order $\alpha-s^{2}$ correction to the Drell-Yan $K$ factor, Nucl. Phys. B, 359, 343--405 (1991) [Erratum: Nucl. Phys. B, 644, 403--404 (2002)]
\bibitem{Melnikov:2006kv}
K. Melnikov and F. Petriello, Electroweak gauge boson production at hadron colliders through $O(\alpha_s^2)$, Phys. Rev. D, 74, 114017 (2006)
\bibitem{Catani:2009sm}
S. Catani, L. Cieri, G. Ferrera, D. de Florian, and M. Grazzini, Vector boson production at hadron colliders: a fully exclusive QCD calculation at NNLO, Phys. Rev. Lett., 103, 082001 (2009)

\bibitem{CMS:2014jvv}
V. Khachatryan et al., Search for dark matter, extra dimensions, and unparticles in monojet events in proton\textendash{}proton collisions at $\sqrt{s} = 8$ TeV, Eur. Phys. J. C, 75, 235 (2015)
\bibitem{ATLAS:2017bf}
M. Aaboud et al., Search for dark matter and other new phenomena in events with an energetic jet and large missing transverse momentum using the ATLAS detector, J. High Energy Phys., 01, 126 (2018)
\bibitem{CMS:2017zts}
A. M. Sirunyan et al., Search for new physics in final states with an energetic jet or a hadronically decaying $W$ or $Z$ boson and transverse momentum imbalance at $\sqrt{s} $ = 13 TeV, Phys. Rev. D, 97, 092005 (2018)
\bibitem{ATLAS:2018nda}
M. Aaboud et al., Search for dark matter in events with a hadronically decaying vector boson and missing transverse momentum in $pp$ collisions at $\sqrt{s} = 13$ TeV with the ATLAS detector, J. High Energy Phys., 10, 180 (2018)
\bibitem{CMS:2019zmd}
A. M. Sirunyan et al., Search for supersymmetry in proton-proton collisions at 13 TeV in final states with jets and missing transverse momentum, J. High Energy Phys., 10, 244 (2019)
\bibitem{ATLAS:2021kxv}
G. Aad et al., Search for new phenomena in events with an energetic jet and missing transverse momentum in $pp$ collisions at $\sqrt {s}$ =13  TeV with the ATLAS detector, Phys. Rev. D, 103, 112006 (2021)
\bibitem{CMS:2021far}
A. Tumasyen et al., Search for new particles in events with energetic jets and large missing transverse momentum in proton\textendash{}proton collisions at $\sqrt{s} $ = 13 TeV, J. High Energy Phys., 11, 153 (2021)

\bibitem{Abdallah:2015ter}
J. Abdallah et al., Simplified models for dark matter searches at the LHC, Phys. Dark Univ., 9--10, 8--23 (2015)

\bibitem{ATLAS:2018bnv}
M. Aaboud et al., Search for invisible Higgs boson decays in vector boson fusion at $\sqrt{s} = 13$ TeV with the ATLAS detector, Phys. Lett. B, 793, 499--519 (2019)
\bibitem{CMS:2018yfx}
A. M. Sirunyan et al., Search for invisible decays of a Higgs boson produced through vector boson fusion in proton\textendash{}proton collisions at $\sqrt{s} =$ 13 TeV, Phys. Lett. B, 793, 520--551 (2019)

\bibitem{CMS:2020hkd}
A. M. Sirunyan et al., Measurement of the Z boson differential production cross section using its invisible decay mode (Z$\nu\bar{\nu}$) in proton-proton collisions at $\sqrt{s}=$ 13 TeV, J. High Energy Phys., 05, 205 (2021)
\bibitem{ATLAS:2017txd}
M. Aaboud et al., Measurement of detector-corrected observables sensitive to the anomalous production of events with jets and large missing transverse momentum in $pp$ collisions at $\sqrt{s}=13$ TeV using the ATLAS detector, Eur. Phys. J. C, 77, 765 (2017)
\bibitem{ATLAS:2019xh}
M. Aaboud et al., Measurement of $ZZ$ production in the $\ell\ell\nu\nu$ final state with the ATLAS detector in $pp$ collisions at $\sqrt{s} = 13$ TeV, J. High Energy Phys., 10, 127 (2019)

\bibitem{Grazzini:2017mhc}
M. Grazzini, S. Kallweit, and M. Wiesemann, Fully differential NNLO computations with MATRIX, Eur. Phys. J. C, 78, 537 (2018)
\bibitem{Grazzini:2019jkl}
M. Grazzini, S. Kallweit, J. M. Lindert, S. Pozzorini, and M. Wiesemann, NNLO QCD + NLO EW with Matrix+OpenLoops: precise predictions for vector-boson pair production, JHEP, 02, 087 (2020)

\bibitem{Catani:1996jh}
S. Catani and M. H. Seymour, The Dipole formalism for the calculation of QCD jet cross-sections at next-to-leading order, Phys. Lett. B, 378, 287-301 (1996)
\bibitem{Catani:1996vz}
S. Catani and M. H. Seymour, A General algorithm for calculating jet cross-sections in NLO QCD, Nucl. Phys. B, 485, 291-419 (1997) [Erratum: Nucl. Phys. B, 510, 503-504 (1998)]
\bibitem{Dittmaier:1999mb}
S. Dittmaier, A General approach to photon radiation off fermions, Nucl. Phys. B, 565, 69-122 (2000)
\bibitem{Dittmaier:2008md}
S. Dittmaier, A. Kabelschacht and T. Kasprzik, Polarized QED splittings of massive fermions and dipole subtraction for non-collinear-safe observables, Nucl. Phys. B, 800, 146-189 (2008) 
\bibitem{Gehrmann:2010ry}
T. Gehrmann and N. Greiner, Photon Radiation with MadDipole, J. High Energy Phys., 12, 050 (2010) 050
\bibitem{Kallweit:2017khh}
S. Kallweit, J.M. Lindert, S. Pozzorini and M. Sch\"onherr, NLO QCD+EW predictions for $2\ell2\nu$ diboson signatures at the LHC, J. High Energy Phys., 11, 120 (2017)
\bibitem{Schonherr:2017qcj}
M. Sch\"onherr, An automated subtraction of NLO EW infrared divergences, Eur. Phys. J. C, 78, 119 (2018) 

\bibitem{Catani:2007vq}
S. Catani and M. Grazzini, An NNLO subtraction formalism in hadron collisions and its application to Higgs boson production at the LHC, Phys. Rev. Lett., 98, 222002 (2007)
\bibitem{Catani:2012qa}
S. Catani, L. Cieri, D. de Florian, G. Ferrera, and M. Grazzini, Vector boson production at hadron colliders: hard-collinear coefficients at the NNLO, Eur. Phys. J. C, 72, 2195 (2012)

\bibitem{Matsuura:1988sm}
T. Matsuura, S. C. van der Marck, W. L. van Neerven, The Calculation of the second order soft and virtual contributions to the Drell-Yan cross-cection, Nucl. Phys. B, 319, 570-622 (1989)
\bibitem{Cascioli:2011va}
F. Cascioli, P. Maierh\"ofer, and S. Pozzorini, Scattering amplitudes with Open Loops, Phys. Rev. Lett., 108, 111601 (2012)
\bibitem{Denner:2016kdg}
A. Denner, S. Dittmaier, and L. Hofer, Collier: a fortran-based complex one-loop lIbrary in extended regularizations, Comput. Phys. Commun., 212, 220-238 (2017)
\bibitem{Buccioni:2017yxi}
F. Buccioni, S. Pozzorini, and M. Zoller, On-the-fly reduction of open loops, Eur. Phys. J. C, 78, 70 (2018)
\bibitem{Buccioni:2019sur}
F. Buccioni, J.-N. Lang, J. M. Lindert, P. Maierh\"ofer, S. Pozzorini, H. Zhang, and M. F. Zoller, OpenLoops 2, Eur. Phys. J. C, 79, 866 (2019)

\bibitem{Bertone:2017bme}
V. Bertone, S. Carrazza, N. P. Hartland, and J. Rojo, Illuminating the photon content of the proton within a global PDF analysis, SciPost Phys, 5, 008 (2018)
\bibitem{Manohar:2016nzj}
A. Manohar, P. Nason, G. P. Salam, and G. Zanderighi, How bright is the proton? A precise determination of the photon parton distribution function, Phys. Rev. Lett., 117, 242002 (2016)
\bibitem{Buckley:2014ana}
A. Buckley, J. Ferrando, S. Lloyd, K. Nordstrom, and B. Page, LHAPDF6: parton density access in the LHC precision era, Eur. Phys. J. C, 75, 132 (2015)

\bibitem{Denner:2005fg}
A. Denner, S. Dittmaier, M. Roth, and L. H. Wieders, Electroweak corrections to charged-current e+ e- --\ensuremath{>} 4 fermion processes: Technical details and further results, Nucl. Phys. B, 724, 247-294 (2005) [Erratum: Nucl. Phys. B, 854, 504-507 (2012)]

\bibitem{ParticleDataGroup:2016lqr}
C. Patrignani et al., Review of Particle Physics, Chin. Phys. C, 40, 100001 (2016)

\bibitem{Catani:2010en}
S. Catani, G. Ferrera, and M. Grazzini, W boson production at hadron colliders: the lepton charge asymmetry in NNLO QCD, J. High Energy Phys., 05, 006 (2010) 

\bibitem{CMS:2021ctt}
A. M. Sirunyan et al., Search for resonant and nonresonant new phenomena in high-mass dilepton final states at $ \sqrt{s} $ = 13 TeV, J. High Energy Phys., 07, 208 (2021)
\bibitem{ATLAS:2019erb}
G. Aad et al., Search for high-mass dilepton resonances using 139 fb$^{-1}$ of $pp$ collision data collected at $\sqrt{s}=$13 TeV with the ATLAS detector, Phys. Lett. B, 796, 68--87 (2019)

\end{thebibliography}
\end{document}